\documentclass[onecollarge,natbib]{svjour2}
\bibpunct{[}{]}{;}{n}{}{,}
\smartqed
\usepackage{amsmath}
\usepackage{graphicx}

\journalname{Few-Body Systems (EFB22)}

\begin{document}

\title{The exact solution for the Dirac equation with the Cornell
  potential 
  \thanks{The authors would like to thank L. Tomio for helpfull
    discussions. L.A. Trevisan and F.M. Andrade thanks the Funda\c{c}\~{a}o
    Arauc\'{a}ria and Setor de Ci\^{e}ncias Exatas e Naturais for
    support.}
}
\author{L. A. Trevisan \and 
        C. Mirez \and 
        F. M. Andrade }

\institute{L.A. Trevisan and F.M. Andrade\at
              Departamento de Matem\'atica e Estat\'\i stica,
              Universidade Estadual de Ponta Grossa, 
              84010-790, Ponta Grossa, PR, Brazil \\ 
              \email{luisaugustotrevisan@yahoo.com.br}           
           \and
           C. Mirez \at
           UFVJM, Campus Teofilo Otoni }

\date{Received: date / Accepted: date}

\maketitle

\begin{abstract}
An analytical solution of the Dirac equation with a Cornell potential,
with identical scalar and vectorial parts, is presented.
The solution is obtained by using the linear potential solution, related
to Airy functions, multiplied by another function to be determined.
The energy levels are obtained and we notice that they obey a band
structure.  

\end{abstract}

\section{Introduction}
\label{intro}
The Cornell potential \cite{Ref1,Ref2} was proposed to describe the
interaction of quarks, containing a confining linear term and a Coulomb
interaction.
To our best knowledge, the potential does not possess exact solutions
within the common equations of quantum mechanics, i.e., the
nonrelativistic Schr\"{o}dinger equation, relativistic Dirac,
Klein-Gordon, Proca, and Duffin-Kemmer-Petiau (DKP) equations. 
Here, we focus on  Dirac equation, with equal scalar and vectorial
components.

\section{The Solution for the Cornell Potential}
We consider each quark with mass $m_{\alpha}$ described by a confining
potential $V(r)$ in the Dirac equation, which is given by
\begin{equation}
\left[\vec{\alpha}\cdot\vec{p}+\beta
m_{\alpha}+\frac{1}{2}(1+\beta)V(r)\right]\Psi(\mathbf{x})= 
E_{\alpha n}\Psi(\mathbf{x}),
\end{equation}
where $\vec{\alpha}$ and $\beta$ are the usual Dirac matrices, and the
index $\alpha$ refers to the kind of particle ($u$, $d$, or $s$ quarks,
in the present case).
By decomposing the above equation in spherical coordinates, we have the
component wave function
\begin{equation}
  \Psi(\mathbf{x})=
  \left(
    \begin{array}{c}
      \chi(\mathbf{x}) \\
      \varphi(\mathbf{x})
    \end{array}
  \right)=
  \left(
    \begin{array}{c}
      f(r) \Omega_{\kappa}^{m}\left(\theta,\phi\right) \\
      ig(r) \Omega_{-\kappa}^{m}\left(\theta,\phi\right)
    \end{array}
  \right)
  =\frac{1}{r}
  \left(
    \begin{array}{c}
      u(r) \Omega_{\kappa}^{m}\left(\theta,\phi\right) \\
     iv(r) \Omega_{-\kappa}^{m}\left(\theta,\phi\right)
    \end{array}
  \right),
  \label{eq:wavef}
\end{equation}
such that, after separating the radial parts, we can get the following
two radial equations
\begin{subequations}
  \begin{align}
    \frac{du(r)}{dr} = {} &
    -\frac{\kappa}{r}u(r)+[E_{\alpha n}+m_{\alpha}]v(r),
    \label{eq:radialeqa}\\
    \frac{dv(r)}{dr} = {} &  \frac{\kappa}{r}v(r)-
    [E_{\alpha n}-m_{\alpha}-V(r)]u(r),
    \label{eq:radialeqb}
  \end{align}
  \label{eq:radialeq}%
\end{subequations}
where $\kappa= -\left(j+1/2\right)=-(l+1)$ if
$l=j-1/2$ and $\kappa=j+1/2=l$  if $l=j+1/2$ is
the Dirac kappa. 
The system (\ref{eq:radialeq}) may be reduced to a Schr\"{o}dinger
like equation 
\begin{equation}
\frac{d^2u(r)}{dr^2} -
\frac{\kappa(\kappa+1)}{r^2}u(r)
+\left[E_{\alpha n}+m_{\alpha}\right]
\left[E_{\alpha n}-m_{\alpha}-V(r)\right]u(r)=0.
\label{eq:ueq}
\end{equation}

The solution for the case of a linear potential, $V(r)=\lambda r$, and
$\kappa=-1$ is described below (see details in \cite{Ref3} and
\cite{Ref4}): 
The unnormalized solution of (\ref{eq:ueq}) is $u(r)=
Ai(\lambda^{1/3}r+a_{n})$ where  $Ai(r)$ is the Airy function and
$a_{n}$ are its roots.
The energy levels are given by: 
$E_{\alpha n}= m_{\alpha}-\lambda a_{n}/K_{\alpha}$, where
$K_{\alpha}\equiv \sqrt[3]{\lambda (E_{\alpha n}+m_{\alpha})}$. 
The above form avoids singularities at origin.

For the Cornell potential, $V(r)=\lambda r -\sigma/r$ , we suppose a
solution such as $ u(r)=Ai(\lambda^{1/3}r+a_{n}) F(r)$. 
The energy levels are renamed to $E_{\alpha n'}'+\delta_{n n'}$, with
the $\delta_ {n n'}$ term to be determined in a consistent way with the
energy levels \cite{Ref5}.
The relation between $\delta_{n n'}$ and $E_{\alpha n n'}'$ is :
\begin{equation}
-\frac{(E_{\alpha n n'}'^{2}-m_{\alpha}^{2})}
{\left[(E_{\alpha n n'}'+m_{\alpha}+\delta_{n n'})\lambda\right]^{2/3}}= a_{n}.
\label{1au}
\end{equation}
By replacing $u(r)$ in Eq. (\ref{eq:radialeq}), we notice that it
can be separated in a part that is the Airy equation, and another that,
in the limit $r \to 0$, has the same form of that of the hydrogen atom
(the part where the potential $-\sigma/r$ dominates). 
Therefore, the solution has a known form: $F(r)=P_{n'}(r)e^{-\eta r}
r^{l}$ and (if $l=0$) 
\begin{subequations}
  \begin{align}
    \eta^{2}+\delta_{n n'}(2E_{\alpha n n'}'+\delta_{n n'})= {} & 0,
    \label{eq:alpha1}\\
    -2\eta(1+n')+(E_{\alpha n n'}'+\delta_{n n'}+m_{\alpha})\sigma = {} & 0,
    \label{eq:alpha2}%
  \end{align}
  \label{eq:alpha}%
\end{subequations}
where $n'$ is the quantum number for the resulting equation for $F(r)$.

\section{Results and Conclusion}
Note that there are two quantum numbers, $n$ and $n'$, due to the
relation among equations (\ref{1au}) and (\ref{eq:alpha}). 
In this manner, the energy spectrum has a band structure, and if
$n' \to \infty$, $E_{\alpha n n'}'+\delta_{n n'} \to E_{\alpha n}$. 
The Table \ref{tab:1} shows the numerical calculated energy values for 
the linear potential and also compares with the numerical energy values
for the Cornell potential obtained thorough the first-order perturbation
correction taking the potential $-\sigma/r$ as a perturbation.
The results shown are for $\lambda=400MeV/fm$ and $\sigma=100MeV.fm$
and $m_{\alpha}=0$. 
A more detailed work \cite{Ref5} will be presented soon.

\begin{table}[t]
\caption{The energy levels (MeV), $E_{\alpha n}$ is for the linear
  potential only, $E_{pert}$ is calculated by first order perturbation
  method, $E_{\alpha n n'}'+\delta_{n n'}$ is for the case of the Cornell potential
  with different $n'$.}  
\centering
\label{tab:1}
\begin{tabular}{llllll}
\hline\noalign{\smallskip}
State (n) & $E_{\alpha n}$ & $E_{pert}$ & $E_{\alpha n'}' + \delta_{n n'}$
($n'=0$) & $E_{\alpha n'}' + \delta_{n n'}$ ($n'=1$)& $E_{\alpha n n'}'+\delta$
($n'=2$)\\[3pt] 
\tableheadseprule\noalign{\smallskip}
1 & 531.20 &463.54 & 457.33  &512.21 & 522.71 \\
2 & 807.72 & 769.02  & 695.39 &778.84 & 794.81 \\
3 & 1011.90 & 983.25 & 871.18 &975.72 & 995.74 \\
\noalign{\smallskip}\hline
\end{tabular}
\end{table}


\begin{thebibliography}{3}

\bibitem{Ref1} Eichten E., Gottfried K., Kinoshita T.,   Lane K.D., and
  Yan T.M.:  Phys. Rev. D \textbf{17}, 3090 (1978).
\bibitem{Ref2} Eichten E., Gottfried K., Kinoshita T.,
  Lane K.D., and  Yan T.M.:
  Phys. Rev. D \textbf{21}, 313(E) (1980).
\bibitem{Ref3} Ferreira P.L., Alcaras J.A.C.:
  Lett. A. Nouvo  Cim. \textbf{14} 500 (1975). 
\bibitem{Ref4} Ferreira P.L., Helayel J.A, Zagury N.:
  Nouvo Cim. A \textbf{55}  215 (1980) .
\bibitem{Ref5} Trevisan, L.A., Mirez, C., Andrade, F.M.: To be submitted.

\end{thebibliography}
\end{document}